\documentclass[twocolumn,numberedappendix,iop]{openjournal}
\usepackage{graphicx,amsmath,amssymb,amstext}
\usepackage{amsbsy,amsfonts,amsthm,color}
\usepackage[colorlinks,linkcolor=blue,citecolor=blue,urlcolor=blue ]{hyperref}
\usepackage[utf8]{inputenc}
\usepackage{float}
\usepackage{placeins} 
\usepackage{xcolor}
\usepackage{ulem}
\usepackage[T1]{fontenc}
\usepackage[title]{appendix}
\usepackage{array}
\usepackage{multirow}
\usepackage{comment}

\begin{document}

\title{The role of peculiar velocity uncertainties in standard siren cosmology \vspace{-4em}}

\author{Chris Blake$^{1,2,*}$}
\author{Ryan Turner$^{1,2}$}
\thanks{$^*$E-mail: cblake@swin.edu.au}

\affiliation{$^1$ Centre for Astrophysics and Supercomputing, Swinburne University of Technology, P.O. Box 218, Hawthorn, VIC 3122, Australia}
\affiliation{$^2$ OzGrav: The ARC Centre of Excellence for Gravitational Wave Discovery}

\begin{abstract}
Local distance indicators such as standard sirens, in combination with spectroscopic redshift measurements of their host galaxies, allow us to estimate the present-day expansion rate of the Universe parameterised by Hubble's constant, $H_0$.  However, these observed redshifts are systematically modified by the effect of galaxy peculiar velocities.  Although these velocities may be estimated from the local density field by the process of velocity-field reconstruction, the intrinsic errors and covariance in these estimates contribute to the error in the $H_0$ determination.  In this paper we demonstrate how the impact of peculiar velocities can be propagated into $H_0$ measurements from local distance indicators with observed redshifts, incorporating the full covariance of the velocity field induced by bulk flows.  We apply our methods to cosmological simulations, testing the importance of this effect in the context of future analyses of gravitational wave sources with electromagnetic counterparts used as bright sirens.  We conclude that $H_0$ errors may be increased by a factor of 2 in comparison with neglecting peculiar velocity covariance, for GW170817-like sirens located within 50 Mpc with $\sim 5\%$ distance errors, with the highest impacts expected for sources at nearby distances or with small distance errors.  Our analytical methods may also be applied to other local distance indicators, such as Type Ia supernovae.
\\[1em]
\end{abstract}

\maketitle

\section{Introduction}

The present-day expansion rate of the Universe, known as Hubble's constant $H_0$, is a key parameter of our cosmological model, linking distance and redshift in the local Universe.  Its value may be measured by combining the recession velocity of nearby objects with redshift-independent distance indicators, such as standard candles.  However, current determinations of $H_0$ from local ``distance ladder'' techniques \citep[e.g.,][]{2016ApJ...826...56R, 2019ApJ...882...34F, 2020ApJ...891L...1P, 2021ApJ...911...65B, 2022ApJ...934L...7R, 2025MNRAS.539.3627S, 2025ApJ...985..203F} lie in moderate-to-significant disagreement with predictions of the value of $H_0$ based on the standard cosmological model, calibrated by observations of the Cosmic Microwave Background radiation \citep{2020A&A...641A...6P, 2023PhRvD.108b3510B, 2024ApJ...962..113M} and Baryon Acoustic Oscillations \citep{2011MNRAS.416.3017B, 2021PhRvD.103h3533A, 2025JCAP...02..021A}.  This disagreement is known as the ``Hubble tension'', and has motivated a wide range of potential solutions \citep[see reviews by, e.g.,][]{2018JCAP...09..025M, 2021CQGra..38o3001D, 2021ApJ...919...16F, 2023Univ....9..393V}, many of which challenge the assumptions of the current cosmological model.

Gravitational-wave astronomy has enabled a new type of distance indicator for mapping out local expansion: standard sirens, which result from mergers of compact binary systems of black holes or neutron stars.  By fitting binary coalescence models to the observed gravitational-wave signals, the physical parameters of each merger may be determined, and hence the amplitude of the signal can be used to infer the distance to the event.  If the gravitational-wave source has a known electromagnetic counterpart (``bright siren''), whose redshift may also be measured, the Hubble constant may be directly determined.  For example, the first bright siren event, GW170817, produced a constraint $H_0 = 70^{+12}_{-8}$ km s$^{-1}$ Mpc$^{-1}$ \citep{2017Natur.551...85A}, which has been refined by several later studies \citep[e.g.,][]{2019NatAs...3..940H, 2020MNRAS.492.3803H, 2020MNRAS.495...90N, 2024PhRvD.109f3508P, KellyPaper}.  In the absence of electromagnetic counterparts, gravitational-wave sources may be associated with the distribution of galaxies in a statistical sense as ``dark sirens'' \citep{1986Natur.323..310S, 2019ApJ...876L...7S, 2020ApJ...900L..33P, 2023JCAP...12..023G, 2023AJ....166...22G, 2024MNRAS.528.3249A, 2025arXiv250217747C}.  Whilst the accuracy of standard siren measurements is not currently competitive with other late-Universe determinations of $H_0$, the number of available systems, and the precision of the resulting distance measurements, will improve over coming years, with forecasts presented by, e.g., \cite{2018Natur.562..545C, 2019PhRvD.100j3523M, 2019PhRvL.122f1105F}.

A key challenge when determining the Hubble constant from local distance indicators is that the observed redshifts of objects reflect not just the overall cosmological expansion, but also the local motions or ``peculiar velocities'' of galaxies that result from gravitationally-driven flows towards overdense regions \citep[for a recent review, see][]{2024arXiv241119484T}.  Moreover, the peculiar velocity has the highest fractional contribution to the redshift for the nearby sources which provide the most accurate distance measurements.  As such, peculiar velocities contribute a potentially important source of statistical and systematic error to $H_0$ determinations \citep[see, e.g.,][]{2014PhRvL.112v1301B, 2021PhRvD.104h3506F, 2023MNRAS.526..337T}.  The connection between peculiar velocities and standard siren inferences has been considered by various studies \citep{2018PhRvD..98f3503W, 2019PhRvD.100b3527H, 2020MNRAS.492.3803H, 2020MNRAS.495...90N, 2021A&A...646A..65M, 2024MNRAS.527.2152N}.  As a simple measure, a typical peculiar velocity error range might be marginalized.  Alternatively, the velocity field at the locations of the distance indicators may be ``reconstructed'' from the overall galaxy distribution using linear-theory models or a variety of algorithms \citep[e.g.,][]{1993ApJ...412....1D, 1994ApJ...421L...1N, 1999ApJ...520..413Z, 2012MNRAS.427L..35K, 2015MNRAS.450..317C, 2016MNRAS.457..172L, 2019A&A...625A..64J, 2021MNRAS.507.1557L, 2023JCAP...06..062Q, 2025arXiv250200121S}, and used to correct the observed redshifts, albeit with some margin of error.

An issue generally neglected in previous studies is that both the underlying and reconstructed peculiar velocities of objects are correlated between different locations \citep{1988ApJ...332L...7G, 2024OJAp....7E..87B}, because the peculiar velocity field is organised in large-scale ``bulk flows'' on $\sim 100$ Mpc scales.  In this paper we investigate the impact of these velocity correlations on determinations of $H_0$ from local distance indicators, in the context of simulated bright siren analyses.  We assess the extent to which these correlated velocity errors affect the error in the $H_0$ determination as a function of the number of bright sirens, their distance range, the error in the inferred distances and the inclusion of velocity reconstruction.  Our study hence extends the analysis of \cite{2023MNRAS.526..337T} to include velocity covariance, and should provide a useful starting point for the treatment of correlated peculiar velocity errors in future analyses of standard sirens and other distance indicators.

Our paper is structured as follows: in Sec.~\ref{sec:pv}, we present the statistical model for the underlying and reconstructed peculiar velocities that we adopt in our analysis.  In Sec.~\ref{sec:h0error}, we discuss the propagation of correlated errors from the velocity field to Hubble's constant.  In Sec.~\ref{sec:sim}, we describe the simulated standard siren datasets we create to test our analysis approach, and in Sec.~\ref{sec:results}, we present our results for how velocity correlations impact the errors in $H_0$ as a function of the survey characteristics.  We conclude our work in Sec.~\ref{sec:conc}.

\section{Velocity-field statistics}
\label{sec:pv}

In standard-siren cosmology, we wish to infer the value of Hubble's constant $H_0$ from determinations of the distance $D$ and redshift $z$ of sources.  For the purposes of our analysis and clarity of presentation, we assume a low-redshift approximation in which $D = cz_{\rm cos}/H_0$.  Whilst we caution that this approximation is not appropriate for analysis of real datasets \citep[as discussed by][]{2019MNRAS.490.2948D}, this model is applied self-consistently in our simulated surveys described below, and therefore does not limit this work.  The observed redshift $z$ is a combination of the cosmological redshift $z_{\rm cos}$ and the radial peculiar velocity $\mathrm{u}$,
\begin{equation}
    1 + z = \left( 1 + z_{\rm cos} \right) \, \left( 1 + \frac{\mathrm{u}}{c} \right) .
\label{eq:redshifts}
\end{equation}
In this way, the determination of $H_0$ becomes entangled with our knowledge of the peculiar velocity field, and its errors.

The peculiar velocity field may be estimated from the observed galaxy density field by a process of ``velocity-field reconstruction'', and then applied in Eq.~\ref{eq:redshifts} to ``correct'' the observed redshifts and improve the determination of $H_0$.  However, the reconstructed velocity field is subject to error due to sample variance, noise (from the limited number density of galaxies) and non-linear effects.  \cite{2023MNRAS.526..337T} argued that a useful statistical model for the joint probability of the true underlying velocity $\mathrm{u}(\mathbf{x})$ and the reconstructed velocity $\mathrm{v}(\mathbf{x})$ at a location $\mathbf{x}$ is a correlated Gaussian likelihood distribution,
\begin{equation}
  P(\mathrm{u},\mathrm{v}) \propto \exp{\left[ - \frac{1}{2(1-r^2)} \left( \frac{\mathrm{u}^2}{\sigma_\mathrm{u}^2} - \frac{2 r \mathrm{u} \mathrm{v}}{\sigma_\mathrm{u} \sigma_\mathrm{v}} + \frac{\mathrm{v}^2}{\sigma_\mathrm{v}^2} \right) \right]} ,
\end{equation}
in terms of variances $\sigma_\mathrm{v}^2 = \langle \mathrm{v}^2(\mathbf{x}) \rangle$, $\sigma_\mathrm{u}^2 = \langle \mathrm{u}^2(\mathbf{x}) \rangle$, and the cross-correlation coefficient $r = \sigma_{\mathrm{u} \mathrm{v}}^2/(\sigma_\mathrm{u} \, \sigma_\mathrm{v})$, where $\sigma_{\mathrm{u} \mathrm{v}}^2 = \langle \mathrm{u}(\mathbf{x}) \, \mathrm{v}(\mathbf{x}) \rangle$.  Expressions for $\sigma_\mathrm{u}^2$, $\sigma_\mathrm{v}^2$ and $\sigma_{\mathrm{u} \mathrm{v}}^2$ are given by \cite{2024OJAp....7E..87B} for different types of velocity-field reconstruction.  In this statistical model, given the value of the reconstructed velocity $\mathrm{v}(\mathbf{x})$ at a point, the posterior distribution of the underlying velocity $\mathrm{u}(\mathbf{x})$ is a Gaussian with mean $(r \, \sigma_\mathrm{u} / \sigma_\mathrm{v}) \, \mathrm{v}$ and variance $(1 - r^2) \, \sigma_u^2$.

The values of the underlying and reconstructed velocities are correlated between different locations, because the velocity field is arranged in ``bulk flows'' on $\sim 100$ Mpc scales.  The form of these covariances depends on the type of velocity-field reconstruction applied \citep{2024OJAp....7E..87B}, and will be discussed further in Sec.~\ref{sec:sim}.  These correlations have potentially important implications for joint analyses of multiple standard sirens.

\section{Determining the $H_0$ posterior}
\label{sec:h0error}

We now discuss the propagation of errors from the velocity field to Hubble's constant, $H_0$.  We start by considering a single standard siren, with distance probability distribution $P(D)$ and precisely-measured observed redshift $z$ (throughout this analysis we neglect spectroscopic redshift errors, which are generally negligible).  In the following, we will assume the low-redshift approximation of Eq.~\ref{eq:redshifts},
\begin{equation}
c \, z = H_0 \, D + \mathrm{u} ,
\label{eq:zapprox}
\end{equation}
noting again that this approximation, whilst not appropriate for real datasets, is applied self-consistently in our mock datasets.  For trial values of $H_0$ and $D$, the corresponding value of the peculiar velocity $\mathrm{u}$ may be inferred and compared with the prediction of the reconstructed velocity field $\mathrm{v}$ at the location of the source, generating a likelihood for these values of $H_0$ and $D$.  The posterior probability distribution of $H_0$ is then given by marginalising over $D$,
\begin{equation}
    P(H_0) \propto \int dD \, P(D) \, P(\mathrm{u} \, | \, H_0, z, D, \mathrm{v}) ,
\label{eq:h0ind}
\end{equation}
where, following the mean and variance of the reconstructed velocity field specified in Sec.~\ref{sec:pv},
\begin{equation}
\begin{split}
&P(\mathrm{u} \, | \, H_0, z, D, \mathrm{v}) \propto \\ &\exp{ \left[ - \frac{ \left( c \, z - H_0 \, D - (r \, \sigma_\mathrm{u} / \sigma_\mathrm{v}) \, \mathrm{v} \right)^2 }{(1 - r^2) \, \sigma_u^2} \right] } .
\end{split}
\end{equation}

The $H_0$ posterior for a joint analysis of $N$ standard sirens is not equal to the product of the posteriors for individual sources, because the velocity errors are correlated between positions.  Instead, we must evaluate,
\begin{equation}
\begin{split}
    P(H_0) \propto \int d^N\mathbf{D} \, & P(D_1) \cdots P(D_N) \\ & \times P(\mathrm{u}_\alpha \, | \, H_0, z_\alpha, D_\alpha, \mathrm{v}_\alpha) ,
\end{split}
\label{eq:h0corr}
\end{equation}
where we have introduced $\alpha = [ 1,2,\dots,N ]$ as an index labelling the standard sirens, and the velocity likelihood function is,
\begin{equation}
    P(\mathrm{u}_\alpha \, | \, H_0, z_\alpha, D_\alpha, \mathrm{v}_\alpha ) \propto \exp{\left[ - \frac{1}{2} \, w_\alpha \, [C_\mathrm{w}^{-1}]_{\alpha\beta} \, w_\beta \right]} ,
\label{eq:covdiff}
\end{equation}
where we have adopted a summation convention that repeated indices are summed from $1$ to $N$.  The data vector of inferred velocity differences is,
\begin{equation}
    \label{eq:veldiff}
    w_\alpha = c \, z_\alpha - H_0 \, D_\alpha - (r \, \sigma_\mathrm{u} / \sigma_\mathrm{v}) \, \mathrm{v}_\alpha ,
\end{equation}
and the covariance of the velocity differences $\mathbf{C}_\mathrm{w}$ is evaluated using Eq.~32 in \cite{2024OJAp....7E..87B}, and will be described in more detail in Sec.~\ref{sec:sim} for the velocity reconstruction method we adopt in our current study.

Eq.~\ref{eq:h0corr} can always be evaluated by numerical techniques which integrate over the multi-dimensional distance space, for any form of distance prior $P(D)$.  However, if we represent the luminosity distance probability distributions $P(D_\alpha)$ as Gaussian distributions with mean $\mu_{D,\alpha}$ and standard deviation $\sigma_{D,\alpha}$, Eq.~\ref{eq:h0corr} can be conveniently evaluated by analytical marginalisation techniques \citep{2002MNRAS.335.1193B, 2010MNRAS.408..865T}.  Following the notation of \cite{2010MNRAS.408..865T}, we expand the log-likelihood $\mathcal{L} = -2 \ln{P(\mathrm{u}_\alpha)}$ of the inferred peculiar velocity field values as a Taylor series as a function of $\delta D_\alpha = D_\alpha - \mu_{D,\alpha}$,
\begin{equation}
    \mathcal{L} = L_0 + L_\alpha \, \delta D_\alpha + \frac{1}{2} \, L_{\alpha\beta} \, \delta D_\alpha \, \delta D_\beta ,
\end{equation}
where,
\begin{equation}
\begin{split}
    L_0 &= \left. \mathcal{L} \right|_{D_\alpha=\mu_{D,\alpha}} = w_\alpha \, [C_{\mathrm w}^{-1}]_{\alpha\beta} \, w_\beta , \\
    L_\alpha &= \left. \frac{\partial\mathcal{L}}{\partial D_\alpha} \right|_{D_\alpha=\mu_{D,\alpha}} = -2H_0 \, [C_{\mathrm w}^{-1}]_{\alpha\beta} \, w_\beta , \\
    L_{\alpha\beta} &= \left. \frac{\partial^2\mathcal{L}}{\partial D_\alpha \partial D_\beta} \right|_{D_\alpha=\mu_{D,\alpha}} = 2H_0^2 \, [C_{\mathrm w}^{-1}]_{\alpha\beta} .
\end{split}
\end{equation}
The marginalised log-likelihood $\mathcal{L}_{\rm marg} = -2 \ln{L_{\rm marg}}$, after integrating over the Gaussian priors in $D_\alpha$, is then given by Eq.~15 in \cite{2010MNRAS.408..865T},
\begin{equation}
\begin{split}
    \mathcal{L}_{\rm marg} = & L_0 - \frac{1}{2} L_\alpha \left( L_{\alpha\beta} + 2 [C_D^{-1}]_{\alpha\beta} \right)^{-1} L_\beta + \\ & \mathrm{Tr} \left[ \ln{ \left( \delta^K_{\alpha\beta} + \frac{1}{2} [C_D]_{\alpha\gamma} \, L_{\gamma\beta} \right)} \right] ,
\end{split}
\label{eq:marg}
\end{equation}
where $\delta^K_{\alpha\beta}$ is a Kronecker delta symbol and the covariance matrix of the distance errors is,
\begin{equation}
    [C_D]_{\alpha\beta} = \sigma^2_{D,\alpha} \, \delta^K_{\alpha\beta} .
\end{equation}
Finally, the posterior for the Hubble constant is constructed as $P(H_0) \propto \exp{(-\mathcal{L}_{\rm marg}/2)}$.  Application of Eq.~\ref{eq:marg} allows us to analytically perform the multi-dimensional integral given by Eq.~\ref{eq:h0corr}.

\section{Simulation}
\label{sec:sim}

We tested our formalism for $H_0$ recovery by constructing a model bright siren analysis of a simulated dataset.  For this purpose we used the $z=0$ snapshot of the Gigaparsec WiggleZ N-body (GiggleZ) simulation \citep{2015MNRAS.449.1454P}, which is a dark matter-only simulation run in a $1 \, h^{-1}$ Gpc periodic box with a halo mass resolution of $3.0 \times 10^{11} h^{-1} M_{\odot}$.  This simulation was created using the fiducial WMAP-5 cosmological parameters: $\Omega_m = 0.273$, $\Omega_{\Lambda} = 0.727$, $\Omega_b = 0.0456$, $h = 0.705$, $\sigma_8 = 0.812$ and $n_s = 0.960$.

We used the dark matter particle distribution of the simulation to create a mock bright siren analysis, where we summarize the steps as follows:
\begin{itemize}
    \item We randomly sub-sampled $N$ particles with distances $D_{{\rm true},\alpha}$ in the range $D_{\rm min} < D < D_{\rm max}$ from the observer, where $\alpha = \{1 , 2, ..., N \}$, assuming $D_{\rm min} = 20 \, {\rm Mpc}$.
    \item We converted these distances to observed redshifts $z_\alpha$ using Eq.~\ref{eq:zapprox}, including the true radial velocities of the simulation particles, and assuming a fiducial value of Hubble's constant $H_{0,{\rm fid}} = 70 \, {\rm km} \, {\rm s}^{-1} \, {\rm Mpc}^{-1}$.
    \item We considered two options for assigning distance errors to each model siren.  In the first case, we added a Gaussian error to the distances equal to a constant fraction of the true distances, $\sigma_{D,\alpha} = f_D \, D_{{\rm true},\alpha}$, forming the noisy distances, $D_\alpha$.  In the second option, we scaled the error with distance more realistically, by adopting the model of \cite{2019PhRvD.100j3523M}.  Specifically, we scaled the signal-to-noise ratio (SNR) of the GW strain as $\rho_\alpha = \rho_{\rm min} D_{\rm max}/D_{{\rm true},\alpha}$ where $\rho_{\rm min} = 12$, and the resulting error as $\sigma_{D,\alpha} = D_{{\rm true},\alpha}/\rho_\alpha$.  We refer to this second case as ``SNR-scaled distance errors''.
    \item We applied a velocity-field reconstruction algorithm (as described below) to a sub-sample of the particle distribution in the cuboid, to derive a reconstructed model velocity at each object, $v_\alpha$.  Given a real dataset, this reconstruction algorithm would be applied instead to the observed galaxy distribution from a large-scale structure survey of the local Universe; our implementation is conceptually similar for the purposes of this study.
    \item We used this mock dataset $\{ D_\alpha, z_\alpha, \sigma_{D,\alpha}, v_\alpha \}$ to estimate $H_0$, comparing various analysis approaches as discussed below.
\end{itemize}
We repeated this analysis varying the number of bright sirens $N$, their maximum distance $D_{\rm max}$, and the fractional distance error $f_D$ (or, as an alternative, the SNR-scaled errors).

For convenience of this analysis we assume a ``Fourier-based'' linear-theory velocity reconstruction approach, which uses a density-field defined within a Fourier cuboid.  \cite{2024OJAp....7E..87B} contrast the Fourier-based approach with the more computationally-intensive spherical Fourier-Bessel reconstruction within a curved-sky window function, demonstrating in each case how the sample variance and noise propagates into the statistics of the reconstructed velocity field.  However, in our study we focus instead on the propagation of the velocity field statistics into the posterior probability distribution of $H_0$.  In this case, we simply assume a uniform selection function within the box.

We suppose that the matter density field within the cuboid is $\delta_m(\mathbf{x})$, with Fourier transform $\tilde{\delta}_m(\mathbf{k})$.  The components of the underlying velocity field $\mathrm{u}_i(\mathbf{x})$ at $z=0$ are then given by the standard linear-theory relation,
\begin{equation}
  \mathrm{u}_i(\mathbf{x}) = -iH_0f \int \frac{V_\mathrm{box} \, d^3\mathbf{k}}{(2\pi)^3} \, \frac{k_i}{k^2} \, \tilde{\delta}_m(\mathbf{k}) \, e^{-i\mathbf{k} \cdot \mathbf{x}} ,
\label{eq:ui}
\end{equation}
where $f$ is the growth rate of cosmic structure ($f = 0.49$ at $z = 0$ in the fiducial cosmology of our simulation), $V_\mathrm{box}$ is the volume of the enclosing Fourier cuboid, and we neglect redshift-space distortions throughout this analysis.

In this study we focus on the radial velocity of each source relative to an observer at the origin, $\mathrm{u}(\mathbf{x}) = \sum_i \mathrm{u}_i(\mathbf{x}) \, x_i/x$.    The correlation between the radial velocities at two locations is given from Eq.~\ref{eq:ui} by \citep[e.g.,][]{1988ApJ...332L...7G, 2024MNRAS.527..501B},
\begin{equation}
\begin{split}
    \langle \mathrm{u}(\mathbf{x}) \, & \mathrm{u}(\mathbf{y}) \rangle = H_0^2 f^2 \int \frac{dk}{2\pi^2} \, P_m(k) \\ &\times \left[ \frac{j_1(kr)}{kr} \left( \hat{\mathbf{x}} \cdot \hat{\mathbf{y}} \right) - j_2(kr) \left( \hat{\mathbf{x}} \cdot \hat{\mathbf{r}} \right) \left( \hat{\mathbf{y}} \cdot \hat{\mathbf{r}} \right) \right] ,
\end{split}
\label{eq:uucov}
\end{equation}
where $j_\ell$ are the spherical Bessel functions of order $\ell$ and $P_m(k)$ is the matter power spectrum, which we obtain for the GiggleZ fiducial cosmology using the {\tt CAMB} and {\tt halofit} prescriptions \citep{2000ApJ...538..473L, 2003MNRAS.341.1311S, 2012ApJ...761..152T}.

We now select a sub-sample of density tracers within the cuboid with uniform number density $n_0 = 10^{-3} \, h^3 \, {\rm Mpc}^{-3}$, which are used to perform a velocity-field reconstruction by applying Eq.~\ref{eq:ui} to the noisy overdensity field.  (This process is representative of using a large-scale structure catalogue of the local Universe for this purpose.)  The overdensity field may be smoothed by a Gaussian filter with standard deviation $\lambda$, to reduce noise, damping the underlying power spectra of the sample variance and noise by a factor $D^2(k)$, where $D(k) = e^{-k^2 \lambda^2/2}$ is the Fourier transform of the damping kernel.  We apply a smoothing $\lambda = 10 \, h^{-1}$ Mpc in our analysis, although our results are not sensitive to this choice.

Following \cite{2024OJAp....7E..87B}, the covariance of the reconstructed velocities between two positions, $\langle \mathrm{v}(\mathbf{x}) \, \mathrm{v}(\mathbf{y}) \rangle$, is obtained using the same relation given in Eq.~\ref{eq:uucov} whilst making the replacement,
\begin{equation}
    P_m(k) \rightarrow \left( P_m(k) + \frac{1}{n_0} \right) D^2(k) ,
\end{equation}
which incorporates the effect of the shot noise power spectrum due to the sample density, and the smoothing.  The cross-correlation between the underlying velocity field and the reconstructed velocity field, $\langle \mathrm{u}(\mathbf{x}) \, \mathrm{v}(\mathbf{y}) \rangle$ may be determined by making the replacement,
\begin{equation}
    P_m(k) \rightarrow P_m(k) \, D(k) .
\end{equation}
These covariances allow us to determine the covariance of velocity differences $\mathbf{C}_\mathrm{w}$ introduced in Eq.~\ref{eq:covdiff}, using Eq.~32 in \cite{2024OJAp....7E..87B}:
\begin{equation}
\begin{split}
    &[C_\mathrm{w}]_{\alpha\beta} = \langle \mathrm{u}(\mathbf{x}_\alpha) \, \mathrm{u}(\mathbf{x}_\beta) \rangle - R \, \langle \mathrm{v}(\mathbf{x}_\alpha) \, \mathrm{u}(\mathbf{x}_\beta) \rangle \\ &- R \, \langle \mathrm{u}(\mathbf{x}_\alpha) \, \mathrm{v}(\mathbf{x}_\beta) \rangle + R^2 \, \langle \mathrm{v}(\mathbf{x}_\alpha) \, \mathrm{v}(\mathbf{x}_\beta) \rangle + \epsilon_u^2 \, \delta^K_{\alpha\beta}  ,
\end{split}
\label{eq:cov}
\end{equation}
where the factor $R = r \sigma_\mathrm{u} / \sigma_\mathrm{v}$ and $\epsilon_\mathrm{u}$ is the additional noise in the underlying velocities due to, for example, non-linear effects.  We assume $\epsilon_\mathrm{u} = 150 \, {\rm km} \, {\rm s}^{-1}$ for the purposes of this study, which is a standard choice for residual velocity noise \citep{2015MNRAS.450..317C, 2018ApJ...859..101S}.  Hence, we distinguish two contributions to the velocity errors: linear-theory covariance (due to the imperfection of velocity reconstruction, represented by the first four terms of Eq.~\ref{eq:cov}) and non-linear noise (represented by the final term).  For our fiducial assumptions the values of the coefficients are given by $\sigma_u = 337 \, {\rm km} \, {\rm s}^{-1}$, $\sigma_v = 220 \, {\rm km} \, {\rm s}^{-1}$ and $r = 0.72$.

In order to explore the impact of peculiar velocity errors on the determination of $H_0$ we consider three different analysis scenarios, which we outline as follows:
\begin{itemize}
    \item \textbf{Scenario 1:} We determined the posterior probability distribution for $H_0$ for each individual object, including non-linear noise but neglecting linear-theory covariance.  In this case we apply Eq.~\ref{eq:h0ind} for each object, multiply the resulting probability distributions, and hence assume an error budget $[C_\mathrm{w}]_{\alpha\beta} = \epsilon_u^2 \, \delta^K_{\alpha\beta}$ where $\epsilon_\mathrm{u} = 150 \, {\rm km} \, {\rm s}^{-1}$.  This scenario represents the standard treatment of peculiar velocities in current bright-siren analyses.
    \item \textbf{Scenario 2:} We determined the posterior probability distribution for $H_0$ including the full velocity correlations, by applying Eq.~\ref{eq:h0corr} to the dataset.
    \item \textbf{Scenario 3:} We determined the posterior probability distribution for $H_0$ including the full velocity correlations but not using the information from velocity-field reconstruction, i.e., assuming $v_\alpha = 0$ and $[C_\mathrm{w}]_{\alpha\beta} = \langle \mathrm{u}(\mathbf{x}_\alpha) \, \mathrm{u}(\mathbf{x}_\beta) \rangle + \epsilon_u^2 \, \delta^K_{\alpha\beta}$ in Eq.~\ref{eq:cov}.
\end{itemize}
In the following section we compare the results of applying these different analysis approaches for various choices of $N$, the number of sirens, $D_{\rm max}$, the maximum observed distance, and $f_D$, the fractional error in those distance measurements. We considered $N = [2,5,10,25,100]$, $D_{\rm max} = [50, 100, 150, 200]$ Mpc, and $f_{D} = [0.01, 0.02, 0.05,0.10]$ to define this parameter space, adding the SNR-scaled errors as an alternative to $f_D$, making for 100 total tests. We repeated each of these tests for 50 different sub-samples drawn from the N-body simulation, saving the mean and standard deviation obtained from the $H_0$ posterior for each realisation.  We then use the mean of these sets of results for a final measurement for each configuration.

Whilst the focus of our study is not to create realistic GW simulations, our studied configurations span a range of potential future GW datasets.  The ``detection power'' of GW observatories is typically characterised by the distance range for detecting a Binary Neutron Star merger, equivalent to our $D_{\rm max}$ parameter.  This range is currently $\approx 150$ Mpc for LIGO\footnote{\url{https://observing.docs.ligo.org/plan}}, although substantially lower for the Virgo ($\approx 50$ Mpc) and KAGRA ($\approx 10$ Mpc) GW observatories, which may be relevant because the identification of bright sirens with electromagnetic counterparts depends partly on localisation.  The first bright siren event, GW170817, is located at a distance $\approx 44$ Mpc, with an initial distance error of $\approx 15\%$ \citep{2017Natur.551...85A}, which has been refined to $5-10\%$ by follow-up studies \citep[e.g.,][]{2019NatAs...3..940H, 2020MNRAS.492.3803H, 2022Natur.610..273M, KellyPaper}.

\section{Results}
\label{sec:results}

\begin{figure}
    \centering
    \includegraphics[width=\columnwidth]{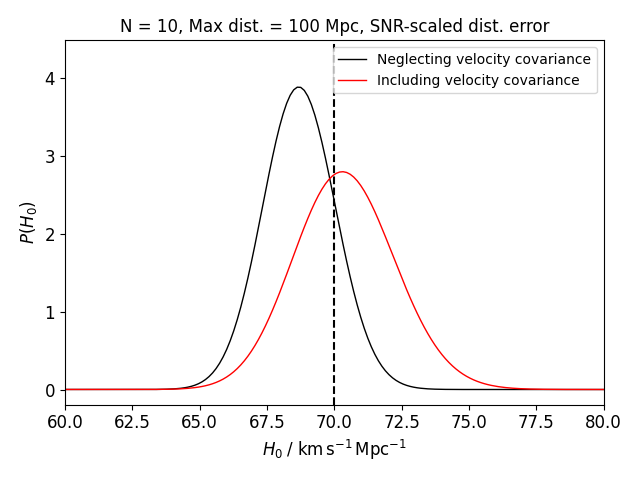}
    \caption{The $H_0$ posterior probability distribution for a bright siren analysis of $N = 10$ sirens distributed to $D_{\rm max} = 100$ Mpc with SNR-scaled distance errors.  The narrower black posterior represents the result of an analysis neglecting velocity covariance (Scenario 1 as presented in Sec.~\ref{sec:sim}), whilst the broader red posterior corresponds to including the full effects of the velocity covariance (Scenario 2).  The difference in the $H_0$ error between these two cases gets larger as the siren distance and distance error reduce.}
    \label{fig:h0posterior}
\end{figure}

As an initial illustration of the effects of fully marginalising over peculiar velocity uncertainties, in Fig.~\ref{fig:h0posterior} we display the $H_0$ posteriors for Scenario 1 (neglecting linear-theory velocity covariance) and Scenario 2 (including velocity covariance) for a single realisation with a fiducial set of survey configuration parameters: $N = 10$ sirens distributed to $D_{\rm max} = 100$ Mpc with SNR-scaled distance errors.  The result indicates that the $H_0$ posterior is broadened by the inclusion of the peculiar velocity covariance, with the error in $H_0$ increasing from $1.2\%$ to $1.8\%$.  Fig.~\ref{fig:h0posterior} also demonstrates that the recovered $H_0$ posteriors are consistent with the input value $H_{0,{\rm fid}} = 70 \, {\rm km} \, {\rm s}^{-1} \, {\rm Mpc}^{-1}$.

\begin{figure}
    \centering
    \includegraphics[width=\columnwidth]{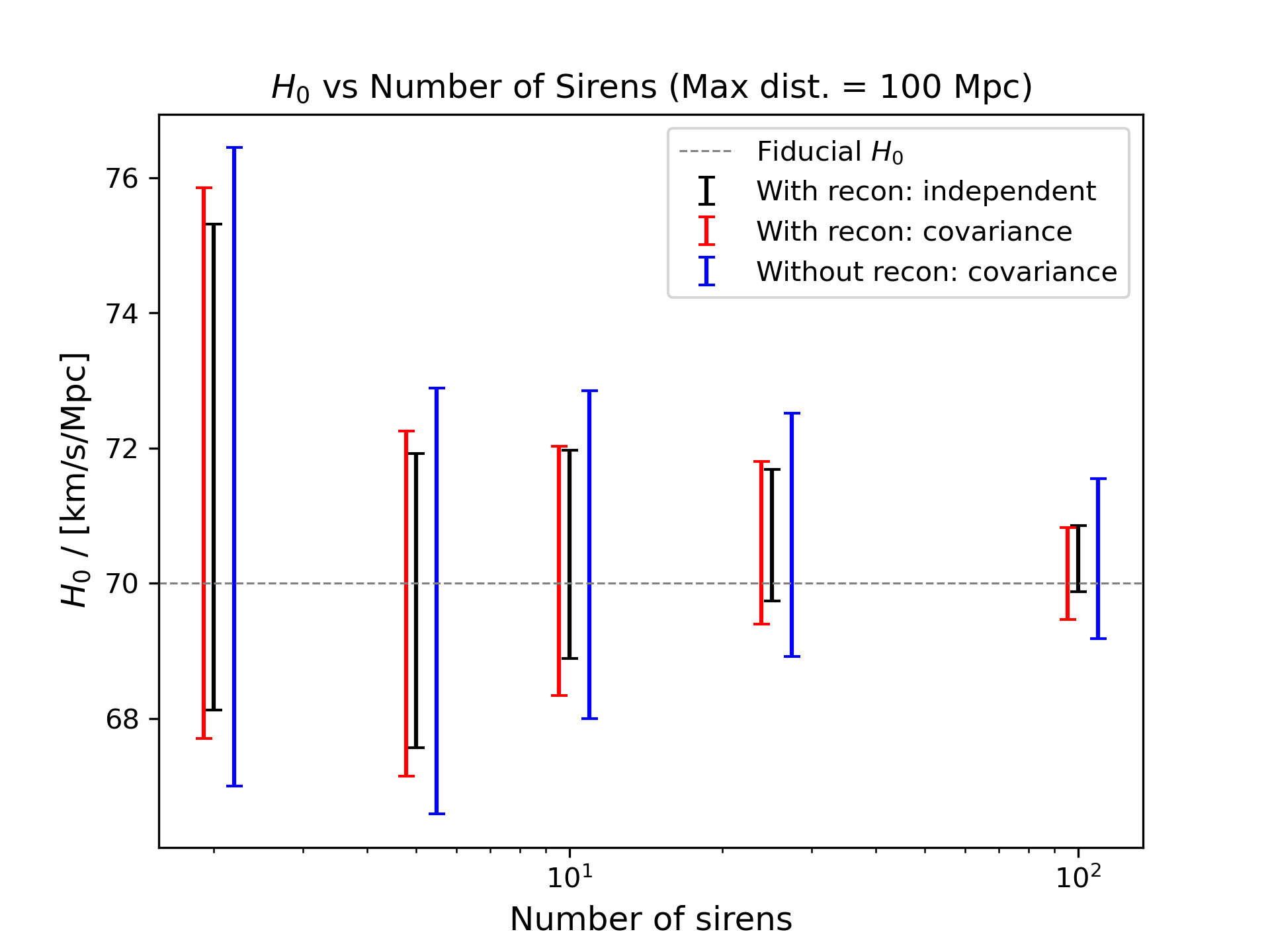}
    \caption{The recovered values of $H_0$ as a function of the number of sirens observed. We fix the maximum distance from the observer $D_{\rm max} = 100$ Mpc, use SNR-scaled distance errors, and measure $H_0$ for $N = [2,5,10,25,100]$.  Results from $H_0$ posteriors derived for a set of independent sirens including reconstruction (Scenario 1) are shown in black, results for correlated sirens including reconstruction (Scenario 2) are shown in red, and results for correlated sirens excluding reconstruction (Scenario 3) are shown in blue.  The fiducial value of $H_0 = 70$ km s$^{-1}$ Mpc$^{-1}$ for the simulation is shown as a dotted line.}
    \label{fig:h0fits}
\end{figure}

Next, we consider results varying one of the survey configuration parameters whilst fixing the remaining two at these fiducial values.  In Fig.~\ref{fig:h0fits} we show the mean recovered value of $H_0$ as a function of the number of sirens included in the analysis, $N$.  Regardless of the number of sirens considered, we are able to recover the fiducial input value of $H_0$.  This is true for both independent and correlated sirens, including and excluding velocity reconstructions.  Hence, we conclude that the inclusion or exclusion of velocity covariance does not significantly bias the recovered value of $H_0$, but rather affects the error in the determined values.  We note that the mean of the recovered $H_0$ values lies above the fiducial value simply because all realisations are drawn from the same simulation, such that there is a residual sample variance which does not average to zero; this offset is unimportant for the conclusions of our study.

\begin{figure}
    \centering
    \includegraphics[width=\columnwidth]{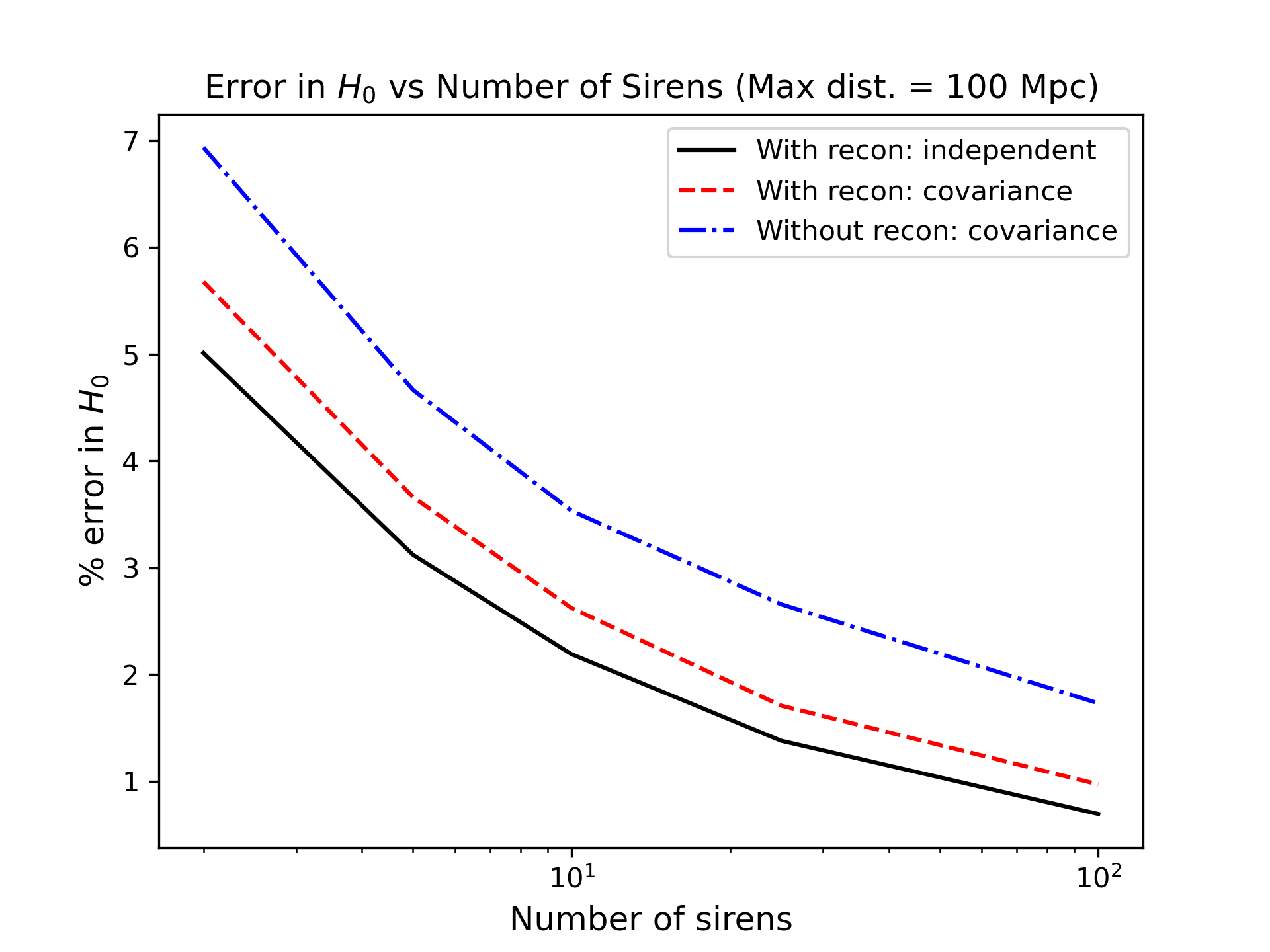}
    \caption{The fractional error in measurements of $H_0$ as a function of the number of sirens observed. We fix the maximum distance from the observer $D_{\rm max} = 100$ Mpc, and use SNR-scaled distance errors, and measure $\sigma_{H_0}$ for $N = [2,5,10,25,100]$. Results from $H_0$ posteriors derived for a set of independent sirens (Scenario 1) are given by the solid black lines, results from sets of correlated sirens including reconstruction (Scenario 2) are given by the dashed red lines, and results for correlated sirens excluding reconstruction (Scenario 3) are given by the dot-dashed blue lines.}
    \label{fig:h0nsiren}
\end{figure}

In Fig.~\ref{fig:h0nsiren}, we display how the $H_0$ measurement accuracy (cast as a percentage) varies as a function of the number of sirens included in the analysis.  Solid lines in black represent the case where we assume that all sirens are independent events and neglect linear velocity covariance (Scenario 1), dashed lines in red represent the case including velocity covariance and incorporating velocity reconstruction (Scenario 2), and dot-dashed lines in blue represent the case including velocity covariance but without incorporating velocity reconstruction (Scenario 3).  We find that including velocity reconstruction information improves the error in $H_0$ by reducing the scatter induced by peculiar velocities.  However, neglecting the covariances in the statistical analysis results in an artificially reduced error in $H_0$.  These differences remain largely the same regardless of the value of $N$.

\begin{figure}
    \centering
    \includegraphics[width=\columnwidth]{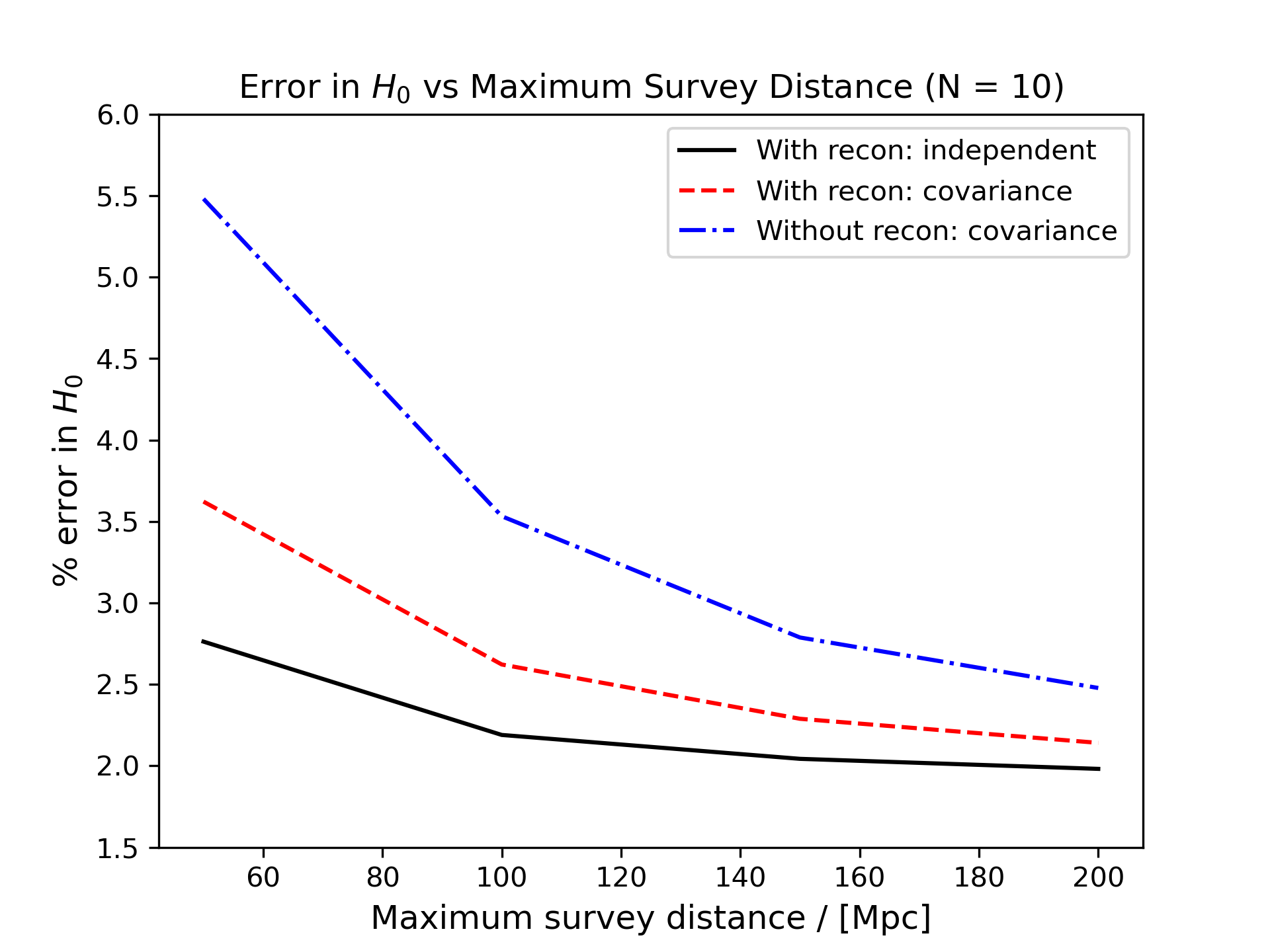}
    \caption{The fractional error in measurements of $H_0$ as a function of the maximum distance from the observer. We fix the number of bright sirens $N = 10$, and use SNR-scaled distance errors, and measure $\sigma_{H_0}$ for $D_{\rm max} = [50, 100, 150, 200]$ Mpc.  The curves are labelled in the same style as Fig.~\ref{fig:h0nsiren}.}
    \label{fig:h0dlmax}
\end{figure}

In Fig.~\ref{fig:h0dlmax}, we show how the $H_0$ measurement accuracy varies as we increase the maximum distance at which sirens are observable, $D_{\rm max}$, fixing the other survey configuration parameters at their fiducial values.  Results are displayed in the same manner as in Fig.~\ref{fig:h0nsiren}.  We note that the importance of velocity correlations increases for sirens located at smaller distances.  This effect arises because velocities have the greatest fractional impact on the observed redshift at low distances, and velocity correlations are more significant over smaller volumes.  As in Fig.~\ref{fig:h0nsiren}, excluding the velocity reconstruction in the analysis results in larger $H_0$ errors for all $D_{\rm max}$.

\begin{figure}
    \centering
    \includegraphics[width=\columnwidth]{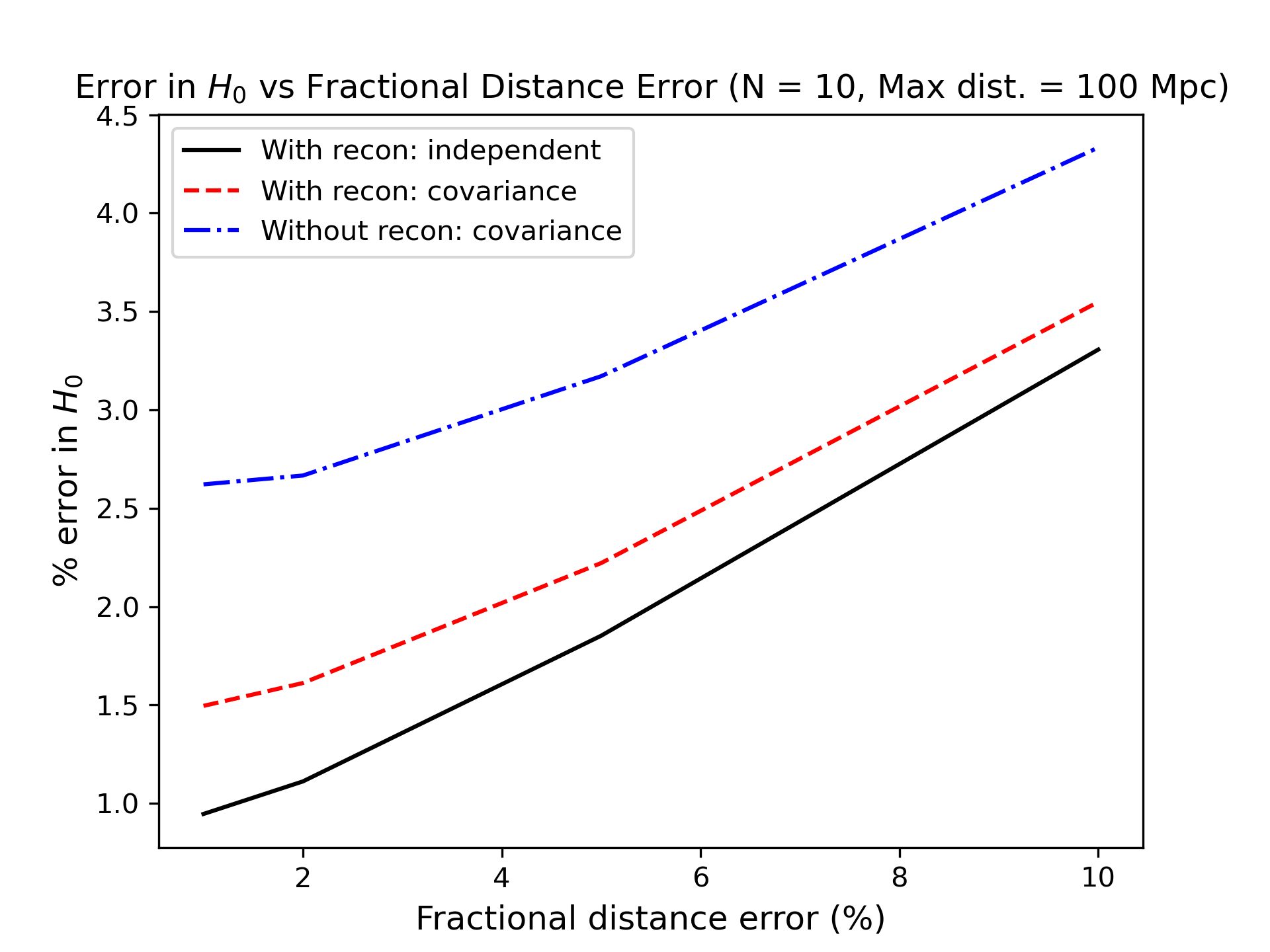}
    \caption{The fractional error in measurements of $H_0$ as a function of the fractional error in the estimate of the host galaxy luminosity distance. We fix the number of bright sirens $N = 10$ and the maximum distance from the observer $D_{\rm max} = 100$ Mpc, and measure $\sigma_{H_0}$ for $f_{D} = [0.01, 0.02, 0.05,0.10]$.  The curves are labelled in the same style as Fig.~\ref{fig:h0nsiren}.}
    \label{fig:h0dlfrac}
\end{figure}

In Fig.~\ref{fig:h0dlfrac} we show how the $H_0$ measurement accuracy behaves as a function of the fractional distance error, $f_D$, fixing the other survey configuration parameters at their fiducial values.  We see that there is an intuitive relationship between the size of the distance error and the size of the error in $H_0$.  The relative difference in results between the cases increases for smaller distance errors, for which peculiar velocities have a larger fractional contribution to the error budget.  As in Fig.~\ref{fig:h0nsiren}, if no reconstructed model velocity is included then the error is $H_0$ is significantly larger for all values of $f_D$.

From our exploration of the survey configuration parameter space we hence find that, whilst the introduction of correlated velocity errors has an impact in all scenarios, the relative effects are greatest for smaller survey distances and smaller distance measurement errors.  Neglecting the information from velocity-field reconstructions  further exacerbates these differences.  The error in $H_0$ resulting from our simulated analyses lies in broad agreement with other bright-siren forecasts \citep[e.g.,][]{2018Natur.562..545C, 2019PhRvD.100j3523M, 2019PhRvL.122f1105F}, when consistent settings are chosen for input assumptions such as the magnitude of distance errors, and their scaling with the signal-to-noise of the input data or distance.  These previous studies neglect the velocity covariance, but perform more sophisticated modelling of the gravitational-wave dataset that is beyond the scope of our analysis.

\section{Conclusions}
\label{sec:conc}

In this study we have presented a self-consistent analysis of the propagation of peculiar velocity effects and uncertainties into the determination of Hubble's constant $H_0$ using local distance indicators with observed host galaxy redshifts.  We apply and validate our method by performing a bright siren analysis drawn from an N-body simulation, focussing on the information gained by applying velocity-field reconstruction to the observed density field to estimate the peculiar velocities of each source.  We summarise our conclusions as follows:
\begin{itemize}
    \item Correcting the observed redshifts in a bright siren analysis, using information from velocity-field reconstruction,  significantly improves the error in the determination of $H_0$, by reducing the scatter imprinted by peculiar velocities.  The improvement in the $H_0$ error by applying reconstruction is around $50\%$ in the scenarios we consider.
    \item However, consistently propagating the errors in velocity-field reconstruction caused by sample variance and noise, and the point-to-point velocity covariance induced by bulk flows, causes the error in $H_0$ to increase appreciably in comparison with neglecting these effects.
    \item The impact of the peculiar velocity covariance is highest for sirens at low distances ($< 100$ Mpc), for which peculiar velocities represent a larger relative contribution to the observed redshift, which is highly correlated by bulk flows.  The impact also increases for sirens with small distance errors, for which peculiar velocity errors represent a larger relative contribution to the overall error budget.  In our simulation studies, the increase in the $H_0$ error approaches a factor of 2 for sirens located within 50 Mpc with $\sim 5\%$ distance errors.
\end{itemize}
Our work may be extended in future by application to real bright or dark siren gravitational wave datasets, together with other sets of local distance indicators such as Type Ia supernovae.

\section*{Acknowledgements}

We thank an anonymous referee for providing useful comments on the original draft of this paper.  This research was conducted by the Australian Research Council Centre of Excellence for Gravitational Wave Discovery (project number CE230100016) and funded by the Australian Government.

\bibliographystyle{mnras}
\bibliography{Bibliography}

\end{document}